\documentclass{aa}
\usepackage{graphics}
\begin{document}

\def\frac{$''$\hspace*{-.1cm}}
\def\deg{$^{\circ}$}
\def\min{$'$}
\def\deg{$^{\circ}$\hspace*{-.1cm}}
\def\min{$'$\hspace*{-.1cm}}
\def\h2{H\,{\sc ii}}
\def\hi{H\,{\sc i}}
\def\hb{H$\beta$}
\def\ha{H$\alpha$}
\def\oiii{[O\,{\sc iii}]}
\def\hei{He\,{\sc i}}
\def\heii{He\,{\sc ii}}
\def\sm{$M_{\odot}$}
\def\lam{$\lambda$}
\def\ab{$\sim$}
\def\x{$\times$}
\def\sec{s$^{-1}$}

\title{
{\it HST} observations of the LMC compact \h2\, region N\,11A\thanks{Based 
   on observations with 
   the NASA/ESA Hubble Space Telescope obtained at the Space Telescope 
   Science Institute, which is operated by the Association of Universities 
   for Research in Astronomy, Inc., under NASA contract NAS\,5-26555.}}

\offprints{M. Heydari-Malayeri, heydari@obspm.fr}

\date{Received  13 March 2001 / Accepted 5 April 2001}

\titlerunning{LMC N\,11A}
\authorrunning{Heydari-Malayeri et al.}

\author{M. Heydari-Malayeri\inst{1} 
        \and 
        V. Charmandaris \inst{2}
        \and 
        L. Deharveng \inst{3}
        \and
        M.\,R. Rosa \inst{4,}\thanks{Affiliated to the Astrophysics
        Division, Space Science Department of the European Space
        Agency.}
        \and   
        D. Schaerer \inst{5} 
        \and  
        H.  Zinnecker \inst{6}}   

\institute{{\sc demirm}, Observatoire de Paris, 61 Avenue de l'Observatoire, 
F-75014 Paris, France 
\and 
Cornell University, Astronomy Department, 106 Space Sciences Bldg.,
Ithaca, NY 14853, USA
\and
Observatoire de Marseille, 2 Place Le Verrier, F-13248 Marseille Cedex
4, France
\and
Space Telescope European Coordinating Facility, European Southern
Observatory, Karl-Schwarzschild-Strasse-2, D-85748 Garching bei
M\"unchen, Germany
\and 
Observatoire Midi-Pyrenees, 14, Avenue E. Belin, F-31400 Toulouse,
France
\and 
Astrophysikalisches Institut Potsdam, An der Sternwarte 16, D-14482
Potsdam, Germany}

\abstract{We present  a study of the LMC compact \h2\, region N\,11A using 
{\it Hubble Space Telescope} imaging observations which resolve N\,11A
and reveal its unknown nebular and stellar features.  The presence of
a sharp ionization front extending over more than 4\frac\, (1 pc) and
fine structure filaments as well as larger loops indicate an
environment typical of massive star formation regions, in agreement
with high \oiii/\hb\, line ratios.  N\,11A is a young region, as deduced
from its morphology, reddening, and especially high local
concentration of dust, as indicated by the Balmer decrement map.  Our
observations also reveal a cluster of stars lying towards the central
part of N\,11A. Five of the stars are packed in an area less than
2\frac\, (0.5 pc), with the most luminous one being a mid O type star. 
N\,11A appears to be the most evolved
compact \h2\, region in the Magellanic Clouds so far studied.
}

\maketitle

\section{Introduction}

The giant \h2\, complex N\,11 (Henize \cite{henize}) or DEM\,34
(Davies et al.\ \cite{dem}) is the most luminous \ha\, source 
in the Large Magellanic Cloud (LMC) after the famous 30 Doradus (Kennicutt
\& Hodge \cite{ken}).  Interestingly, N\,11 has been suggested to be
reminiscent of an evolved, some 2\,\x\,10$^{6}$ years older version of
the 30 Doradus starburst (Walborn \& Parker \cite{wp}).  Known also as
MC\,18 (McGee et al. \cite{mc}), it lies at the north-western
extremity of the LMC bar and consists of a dozen ionized nebulae
(named N\,11A to N\,11L) with several filaments surrounding a central
cavity (see Fig. 1 of Rosado et al.
\cite{ros}). Almost the entire complex is contained
within a  bubble \ab\,25\min\,\x\,20\min\,
(375\,pc\,\x\,300\,pc) in size created by stellar winds and
several supernova explosions (Meaburn et al. \cite{mea}, Rosado et
al. \cite{ros}, Mac Low et al. \cite{mac}, Kim et al. \cite{kim}).
The complex harbors tens of hot stars at different stages of evolution
distributed in four OB associations, LH\,9, LH\,10, LH\,13, and LH\,14 (Lucke
\& Hodge \cite{lh}). \\

A considerable amount of work has been devoted to examining various
properties of the whole complex.  In particular, the physical
properties of the components B (LH\,10), C (LH\,13), 
and D (LH\,14) and their massive star
contents were studied by Heydari-Malayeri \& Testor (\cite{hey83} and
\cite{hey85}, hereafter Papers I and II), Heydari-Malayeri et
al. (\cite{hey87}), Parker et al. (\cite{par}), 
Degioia-Eastwood et al. (1993), 
Heydari-Malayeri \& Beuzit (\cite{hey94}), 
Parker et al. (1996),  
and Heydari-Malayeri et al. (\cite{hey00}). 
Parker  et al. (\cite{par})   present an extensive spectral
classification of some 70 O and B stars in the associations LH\,9 and
LH\,10.  The latter association, which powers N\,11B, is the
largest and brightest component of the complex containing
several young massive stars of type O3. \\

N\,11A (also known as IC\,2116, HD\,32279), to which this paper is
devoted, is a quite isolated ionized region situated north-east of
N\,11B (see Fig. 1 of Paper I). It is a fairly bright object with
$V$\,\ab\,12.5 mag, and its size of \ab\,12\frac\, makes it the
smallest and the most compact nebula in the complex. Using
ground-based narrow-band imaging and spectroscopy, in Papers I and II
we studied several physical characteristics of N\,11A
(emission spectrum, excitation, extinction, gas density, chemical
composition, abundances, etc.), and suggested that it is the most
excited ionized region of the complex. These observations indicated
that the excitation source has a $T_{eff}$\,\ab\,44,000 K, corresponding to
a mid-O type star. Israel \& Koornneef (\cite{ik91}) also presented
infrared and IRAS photometry of N\,11A, and showed that it is a strong
near-IR source dominated by nebular emission. \\

N\,11A belongs to the class of the so-called high excitation blobs
(HEBs) in the Magellanic Clouds. In contrast to the typical \h2
regions of the Magellanic Clouds, which are extended structures spread
over several minutes of arc on the sky and powered by a large number
of hot stars, HEBs are very dense small regions usually 5\frac\, to
10\frac\, in diameter.  At the distance of the Magellanic Clouds this
corresponds to sizes of more than 50 pc for normal \h2\, regions and 1
to 3 pc for the blobs. They are also characterized by large amounts of
local dust.  HEBs are in fact created by very young massive
stars just leaving their natal molecular cloud (see
Heydari-Malayeri et al. \cite{hey1999c} for references). \\

In their study of N\,11A, Parker et al. (\cite{par}) used a method for
removing the contribution of the surrounding nebular brightness in
CCD images and found an embedded star with $V=14.21$ mag and two
fainter neighbors (unpublished images).  The presence of \heii\,
absorption lines in the spectrum of the brightest component (which
they called LH\,10: 3264) confirmed as O the spectral type of
the main exciting source, even though contamination by very strong
nebular emission lines made possible to only restrict its subclass
within the O3--O5\,V range. \\
  
In this paper we use the excellent imaging power of the {\it Hubble
Space Telescope (HST)} to resolve N\,11A and study the structure of the
ionized gas as well as its stellar content.  This allows us to present
a clear picture of the compact ionized nebula and identify its
exciting stars. Such information is essential for a better understanding
of massive star formation, particularly in isolated, compact regions. \\

\begin{figure*}
\begin{center}
\resizebox{16cm}{!}{\includegraphics{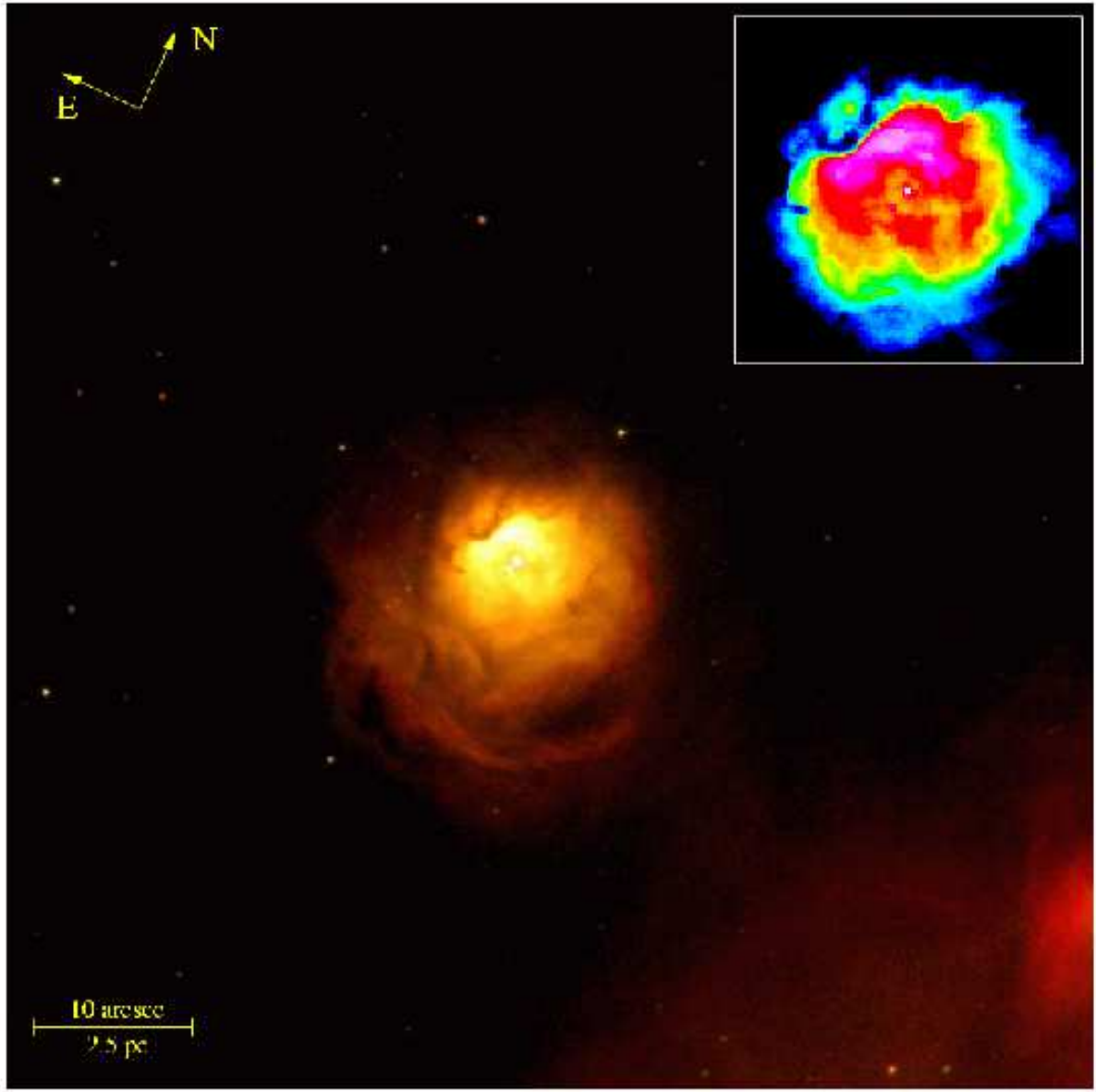}}
\caption{A WFPC2 ``true color'' image of the LMC compact \h2\, region
N\,11A created by combining images taken with the \ha\, (red), \oiii\,
(green), and \hb\, (blue) filters.  The field size is
\ab\,63\frac\,\x\,63\frac\, (\ab\,15\,pc\x\,15\,pc). 
The nebulosity to the south-west belongs to the large
\h2\, region N\,11B. The false-color image shows N\,11A in \ha. 
}
\label{true}
\end{center}
\end{figure*}

\begin{figure*}
\begin{center}
\resizebox{13cm}{!}{\includegraphics{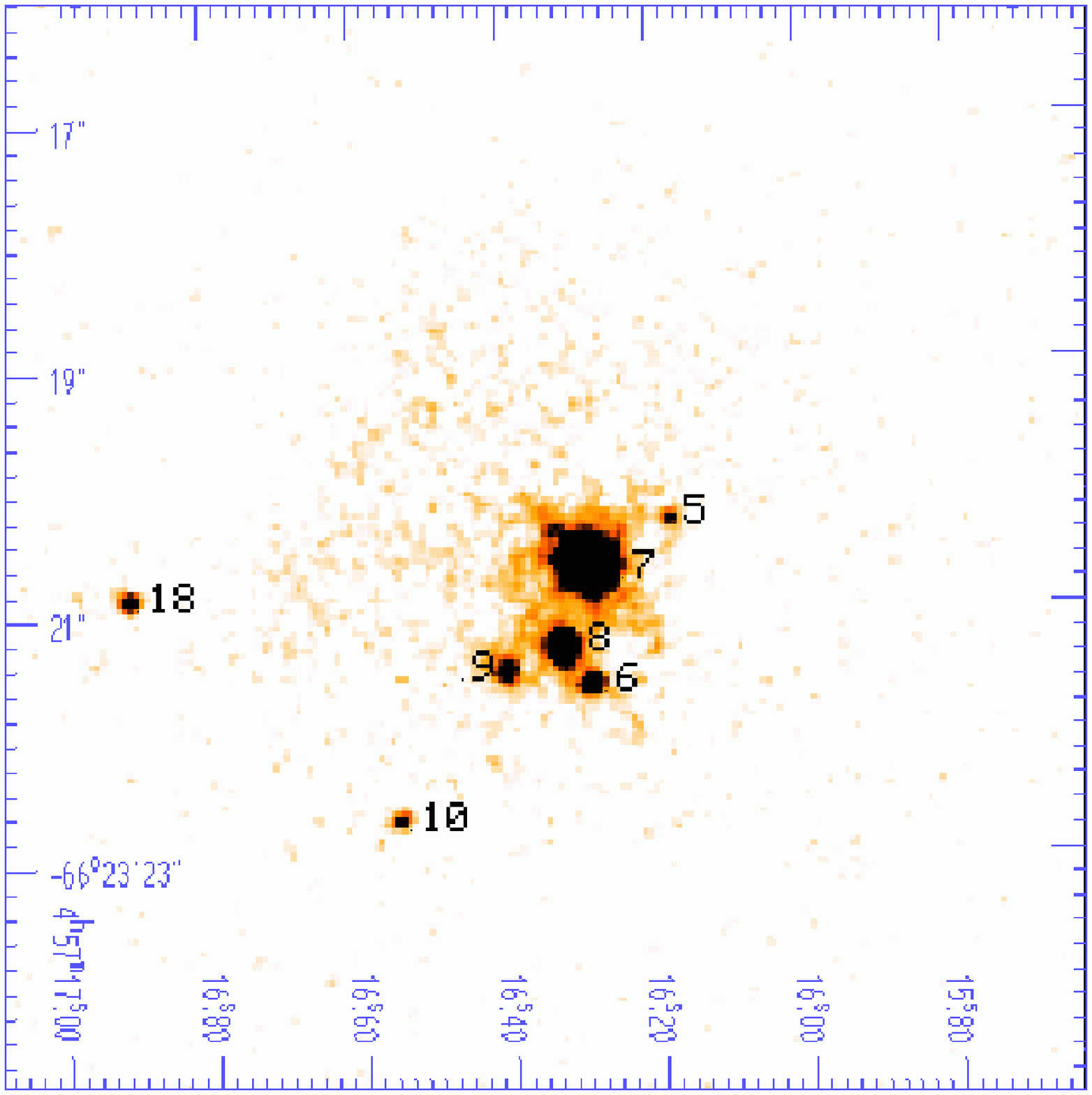}}
\caption{The stellar content of the LMC compact \h2\, region 
N\,11A as revealed by the Str\"omgren $y$ image. The field size 
\ab\,8\frac.8\,\x\,8\frac.8 (\ab\,2.2\,pc\,\x\,2.2\,pc).   
 }
\label{chart_core}
\end{center}
\end{figure*}

\section{Observations and data reduction}

The observations of N\,11A were performed as part of our project GO-8247
with the Wide Field Planetary Camera 2 (WFPC2) on board of the {\it
HST}. The images taken with the broad-band filters (F300W, F467M,
F410M, and F547M) were obtained on February 7, 2000.
Centered on the Planetary Camera (PC), these images 
had as a goal to reveal the details of the stellar content of N\,11A.
The narrow-band filter images
(F487N, F503N and F656N) were obtained on May 21, 2000. In that case
the target was centered on the WF2 which has larger pixels and lower
noise than the PC CCD and is better suited for detecting faint nebular
emission. Exposures were taken at 4 different pointings offset by
0\frac.8 and the exposure times ranged from 10 to 300 sec (see
Table\,\ref{obs} for details). \\

\begin{table}[!h]  
\caption[ ]{N\,11A observations ({\it HST} GO-8247)} 
\label{obs} 
\begin{flushleft}  
\begin{tabular}{lcr}  
\hline 
{\it HST} filter              & Wavelength    & Exposure time\\ 
                        & $\lambda$(\AA)        & (sec)\\
\hline 
F300W (wide-U)                  & 2911          & 4\,\x\,14\,=\,56\\
F410M (Str\"{o}mgren $v$)       & 4090          & 4\,\x\,50\,=\,200\\
F467M (Str\"{o}mgren $b$)       & 4669          & 4\,\x\,35\,=\,140\\
F547M (Str\"{o}mgren $y$)       & 5479          & 4\,\x\,10\,=\,40\\
F487N (H$\beta$)                & 4866          & 4\,\x\,260\,=\,1040\\
F502N ([OIII])                  & 5013          & 4\,\x\,300\,=\,1200\\
F656N  (H$\alpha$)              & 6563          & 4\,\x\,260\,=\,1040\\
\hline    
\end{tabular} 
\end{flushleft}   
\end{table}

The data were processed through the standard {\it HST} pipeline
calibration.  Multiple images were co-added using the {\sc stsdas}
task {\it imcombine}, while cosmic rays were detected and removed with
the {\sc stsdas} task {\it crrej}.  Normalized images were then
created using the total exposure times for each filter.  To extract
the positions of the stars, the routine {\it daofind} was applied to
the images by setting the detection threshold to 5$\sigma$
above the local background level.  The photometry was performed
setting a circular aperture of 3--4 pixels in radius in the {\it
daophot} package in {\sc stsdas}. \\

A crucial point in our data reduction was the sky subtraction. For
most isolated stars the sky level was estimated and subtracted
automatically using an annulus of 6--8 pixel width around each star.
However, this could not be done for several stars located in the
central region of N\,11A due to their crowding. In those cases we
carefully examined the PSF size of each individual star ({\sc
fwhm}\,\ab\,2 pixels, corresponding to 0\frac.09 on the sky) and did
an appropriate sky subtraction using the mean of several nearby
off-star positions.  To convert into a magnitude scale we used zero
points in the Vegamag system, that is, the system where Vega is set to
zero mag in Cousin broad-band filters.  The magnitudes measured were
corrected for geometrical distortion, finite aperture size (Holtzman
et al. \cite{holtz}), and charge transfer efficiency as recommended by
the {\it HST} Data Handbook. An additional correction to take into
account the long versus short photometric anomaly (Casertano \&
Muchler 1998) was also applied. This increased the brightness of our
fainter stars by as much as ~0.4 magnitudes but it had no measurable
effect on the brighter stars of the region.\\

We note that the filter F547M is wider than the standard Str\"omgren
$y$ filter. To evaluate the presence of any systematic effects in our
photometry and color magnitude diagrams due to this difference in the
filters, we used the {\sc stsdas} package {\it synphot}.  Using
synthetic spectra of hot stars, with spectral types similar to those
found in \h2\, regions, we estimated the difference due to the {\it
HST} band-passes to be less than 0.002 mag, which is well within the
photometric errors.  \\

\section{Results}

\subsection{Morphology}

Narrow-band \hb\, and \oiii\, images obtained with ground-based
telescopes show N\,11B to be delineated at its eastern part by a sharp
ionization front (Paper I). In those images N\,11A appears as a
featureless blob situated \ab\,50\frac\, (12 pc) east of the ionization 
front. \\

The present {\it HST} observations resolve N\,11A into a very
bright core surrounded by a diffuse envelope of \ab\,8\frac\, in
diameter (Fig.\,\ref{true}) and reveal several interesting nebular and
stellar features. A sharp ridge or front, extending over \ab\,4\frac\,
along the south-east to north-west direction, borders the core towards
north.  Beyond the ridge the \ha\, emission drops by about a factor of
10.  A small cluster of five tightly packed stars are seen towards a
cavity in the central region (Fig.\,\ref{chart_core}). \\

\begin{figure}
\resizebox{\hsize}{!}{\includegraphics{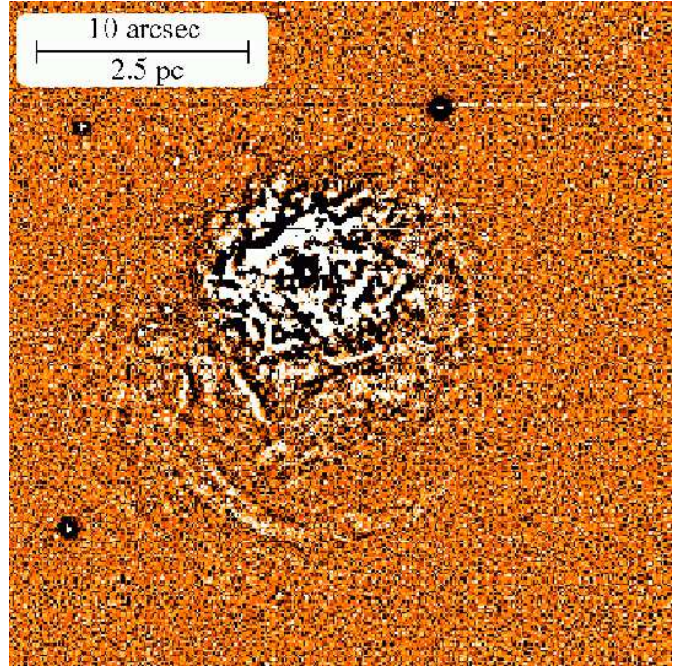}}
\caption{An un-sharp masking image of N\,11A obtained 
in \ha,  which highlights the filamentary  patterns 
of the nebula (see the text). 
The field size is \ab\,30\,\frac\,\x\,30\frac\, (7.5\,pc\,\x\,7.5\,pc), 
and the orientation as in Fig.~\ref{true}}
\label{mask1}
\end{figure}

The small scale filamentary structures are best seen in
Fig.\,\ref{mask1}, which presents an un-sharp masking image of N\,83B in
\ha\, from which large scale structures have been subtracted.  In
order to remove these brightness variations and enhance the high
spatial frequencies, a digital ``mask'' was created from the \ha\,
image.  First, the \ha\, image was convolved by a 2\,\x\,2-pixel
Gaussian, and then the smoothed frame was subtracted from the original
\ha\, image.  \\

\subsection{Extinction and nebular emission}

A map of the \ha/\hb\, Balmer decrement is presented in
Fig.\,4a. It reveals a strong absorption zone north of the
bright ridge, where the \ha/\hb\, ratio reaches values as high as
\ab\,5.5, while its mean value there is 4.5.  Applying the
interstellar reddening law, these ratios correspond to $A_{V}$\,=\,1.8
and 1.3 mag respectively.  Outside that area the ratio is smaller
with a mean value of \ab\,3.5 which corresponds to $A_{V}$\,=\,0.6
mag.  \\

\begin{table*}[t]	
\caption[]{Photometry of the brightest stars towards N\,11A}	
\label{phot}	
\begin{flushleft}
\begin{tabular}{cccccccc} 	
\hline	
Star   &   RA(J2000)     &    Dec(J2000)  &  F300W & F410M & F467M & F547M & Color \\
       &     &        & Wide $U$   &   Str\"omgren $v$ & Str\"omgren $b$ &  Str\"omgren $y$ & $b-y$ \\ 
\hline

    1  &   4:57:13.9  & -66:23:21.0  & 18.74 & 19.30 & 19.47 & 19.29  & +0.18 \\
    2  &   4:57:14.7  & -66:23:07.4  & 18.94 & 19.32 & 19.31 & 19.51  & --0.20\\
    3  &   4:57:15.4  & -66:23:35.7  & 18.38 & 19.36 & 19.42 & 19.40  & +0.02 \\
    4  &   4:57:15.9  & -66:23:10.2  & 14.52 & 16.02 & 16.13 & 16.15  & --0.02\\
    5  &   4:57:16.2  & -66:23:20.3  & --    &   --  &    -- & 20.23  & -- \\
    6  &   4:57:16.3  & -66:23:21.6  & 17.03 & 18.28 & 18.34 & 18.26  & +0.08\\
    7  &   4:57:16.3  & -66:23:20.6  & 12.85 & 14.59 & 14.67 & 14.69  & --0.02\\
    8  &   4:57:16.3  & -66:23:21.3  & 15.16 & 16.70 & 16.67 & 16.68  & --0.01 \\
    9  &   4:57:16.4  & -66:23:21.5  & 17.56 & 18.68 & 18.81 & 18.68  & +0.13\\
   10  &   4:57:16.6  & -66:23:22.7  & 18.13 & 19.92 & 19.27 & 19.41  & --0.14\\
   11  &   4:57:17.2  & -66:23:25.2  & 17.75 & 19.21 & 19.28 & 19.32  & --0.04\\
   12  &   4:57:17.3  & -66:23:26.3  & 18.81 & 19.26 & 19.19 & 19.31  & --0.11\\
   13  &   4:57:17.3  & -66:23:26.8  & 18.22 & 19.15 & 19.13 & 19.14  & -0.01\\
   14  &   4:57:17.4  & -66:23:20.7  & 18.45 & 19.26 & 19.23 & 19.32  & --0.09\\
   15  &   4:57:17.4  & -66:23:09.3  & 18.48 & 19.56 & 19.50 & 19.54  & -0.04\\
   16  &   4:57:17.7  & -66:23:18.3  & 19.18 & 19.31 & 19.34 & 19.38  & -0.04\\
   17  &   4:57:18.5  & -66:23:18.8  & 17.26 & 18.46 & 18.40 & 18.50  & --0.10\\
   18  &   4:57:16.9  & -66:23:20.9  & 17.39 & 19.32 & 18.81 & 18.88  & --0.07  \\
\hline
\end{tabular}
\end{flushleft}
\end{table*}

The [O\,{\sc iii}]$\lambda$5007/\hb\, intensity map
(Fig.\,4b) reveals a remarkably extended high-excitation
zone where the O$^{++}$ ions occupy almost the same volume as H$^{+}$.
The most excited area, though, surrounds the central stellar cluster, 
with  a mean ratio of 4.1, while the ratio reaches slightly 
higher values in some pixels in the zone 
lying between the sharp ridge and the
exciting stars. On  average the \oiii/\hb\,  is \ab\,3.7 over
the diffuse part of the nebula. \\

Using the \ha/\hb\, map, we can accurately correct the \hb\, flux of
the \h2\, region for interstellar reddening on a pixel by pixel basis.  
The total corrected flux is
$F_{0}$(\hb)\,=\,1.17\,$\times$\,10$^{-11}$ erg cm$^{-2}$ s$^{-1}$
above 3$\sigma$ level without the stellar contribution and accurate to
3\,\%, if our extinction correction which assumes external dust is
verified.  Supposing that the \h2\, region is ionization-bounded, the
corresponding Lyman continuum flux of the region is
$N_{L}$\,=\,9.10\,$\times$\,10$^{48}$ photons s$^{-1}$. Similarly,
with the assumption of a homogeneous spherical nebula, a mean electron
density of 550 cm$^{-3}$ can be derived from the \hb\, flux, which
leads to an ionized hydrogen mass estimate of 54\,\sm. \\

A single main sequence star of type O7.5 or O8 or several massive
stars of later types can account for this ionizing photon flux (Vacca et
al. \cite{vacca}, Schaerer \& de Koter \cite{sch}).  However,
this could clearly be an underestimate since the \h2\, region
is more likely to be density-bounded almost in all directions except 
northward where lies the sharp ridge.  \\

\begin{figure*}
\resizebox{\hsize}{!}{\includegraphics{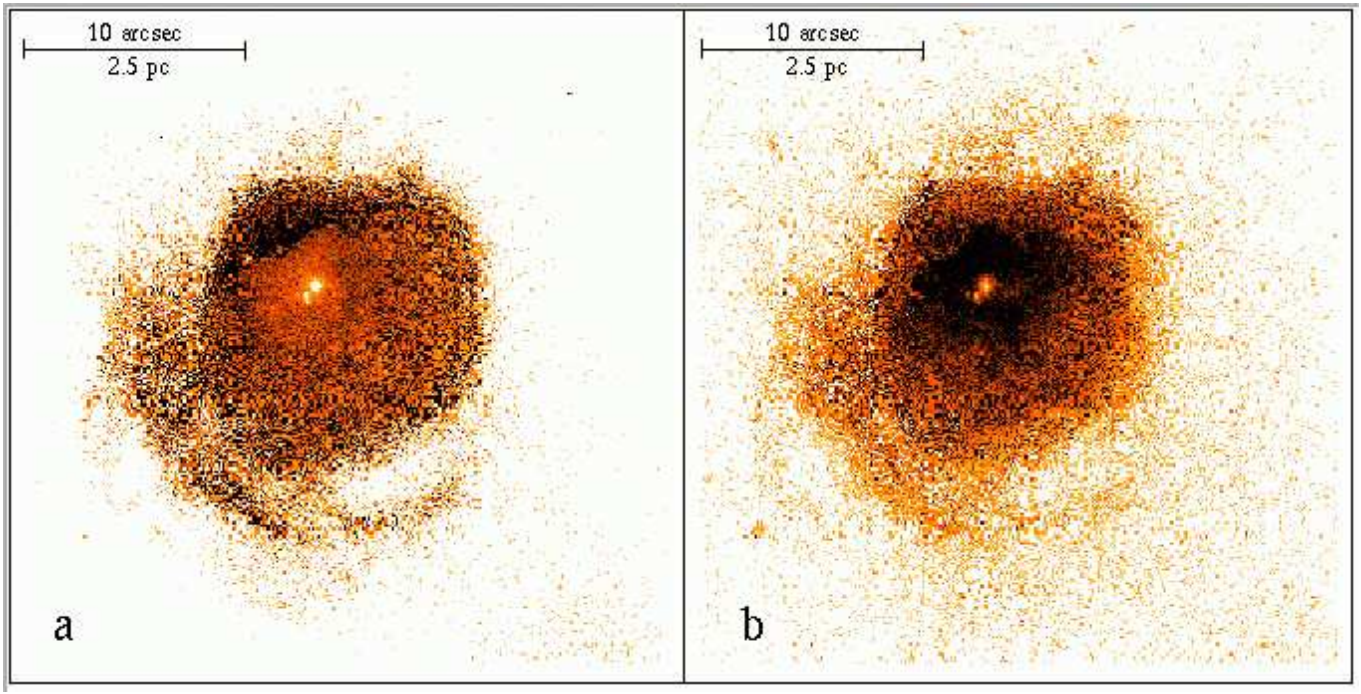}}
\caption{Line intensity ratios for the LMC compact nebula  
N\,11A. Darker colors correspond to higher ratio values.  The field of
view and orientation are identical to Fig.\,\ref{mask1}. The white spots are
stars and can be identified using Fig.\,\ref{chart_core}.  {\bf a)}
\ha/\hb\, Balmer decrement. Its mean value over the diffuse component is
\ab\,3.5 ($A_{V}$\,=\,0.6 mag), while the ratio goes up to \ab\,5.5
($A_{V}$\,=\,1.8 mag) towards the darkest area north-east of the
exciting stars.  {\bf
b)} The [O\,{\sc iii}]$\lambda$5007/\hb\, ratio. The mean value around the 
stars is 4.1, while the highest
values reach \ab\,4.5 in the zone between the exciting stars and the 
sharp ridge. }
\label{ratios}
\end{figure*}

\subsection{Stellar content}

The automatic search routine {\it daofind} detected 18 stars in the
WFPC2 field of N\,11A, all of them unambiguously present in the four
stellar filters (Fig.\,5) with their magnitudes 
ranging from $y=14.69$
to 20.23.  The photometry measurements for these stars are presented
in Table\,\ref{phot}, where the star numbers refer to Figs.\,2 and 4.\\

Among these stars only seven are situated in the direction of N\,11A.  A
small cluster of five stars, observed at the center of N\,11A (Fig.\,2),
is obviously associated with this region.  Star \#7, with a $y$
magnitude of 14.69 is the main exciting star of the region, nearly 6
times more luminous than star
\#8, the second brightest star, which is situated \ab\,0\frac.7 
to its south, and has a magnitude of $y=16.68$. Star \#5
lying 0\frac.8 to the north-east of \#7 is the faintest member of the
sample. If this one is an exciting star, its faint magnitude 
should be due to local reddening. \\

One could try to estimate the luminosity of the brightest star of the
cluster (\#7), although in the absence of spectroscopic data this is
not accurate.  Using an extinction of $A_{V}$\,=\,0.6 mag
corresponding to the mean value for the associated nebula (Section
3.2), and a distance modulus $m$\,--\,$M$\,=\,18.5 (e.g.  Kov\'acs
\cite{ko} and references therein), we find a visual absolute magnitude
$M_{V}=-4.4$.  If the star is on the main sequence, it would be an
O9V according to the calibration of Vacca et al. (\cite{vacca}) for
Galactic stars. Its corresponding luminosity and mass would be {\it
log L}\,=\,5.06\,$L_{\odot}$\, and $M$\,=\,25\,\sm. \\

Parker et al. (\cite{par}) derived a more luminous exciting star with
$M_{V}=-5.1$ mag corresponding to an O6V type. The reason for this
discrepancy is that they measure a $V$ magnitude which is 0.48 mag
brighter than our Str\"omgren $y$ and also use a stronger visual
extinction of $A_{V}$\,=\,0.8 mag. A magnitude offset of this type
could be attributed to the fact that the exciting stars are tightly
packed inside a very bright \h2\, region, making sky subtraction
difficult. \\

\begin{figure*}
\begin{center}
\resizebox{15cm}{!}{\includegraphics{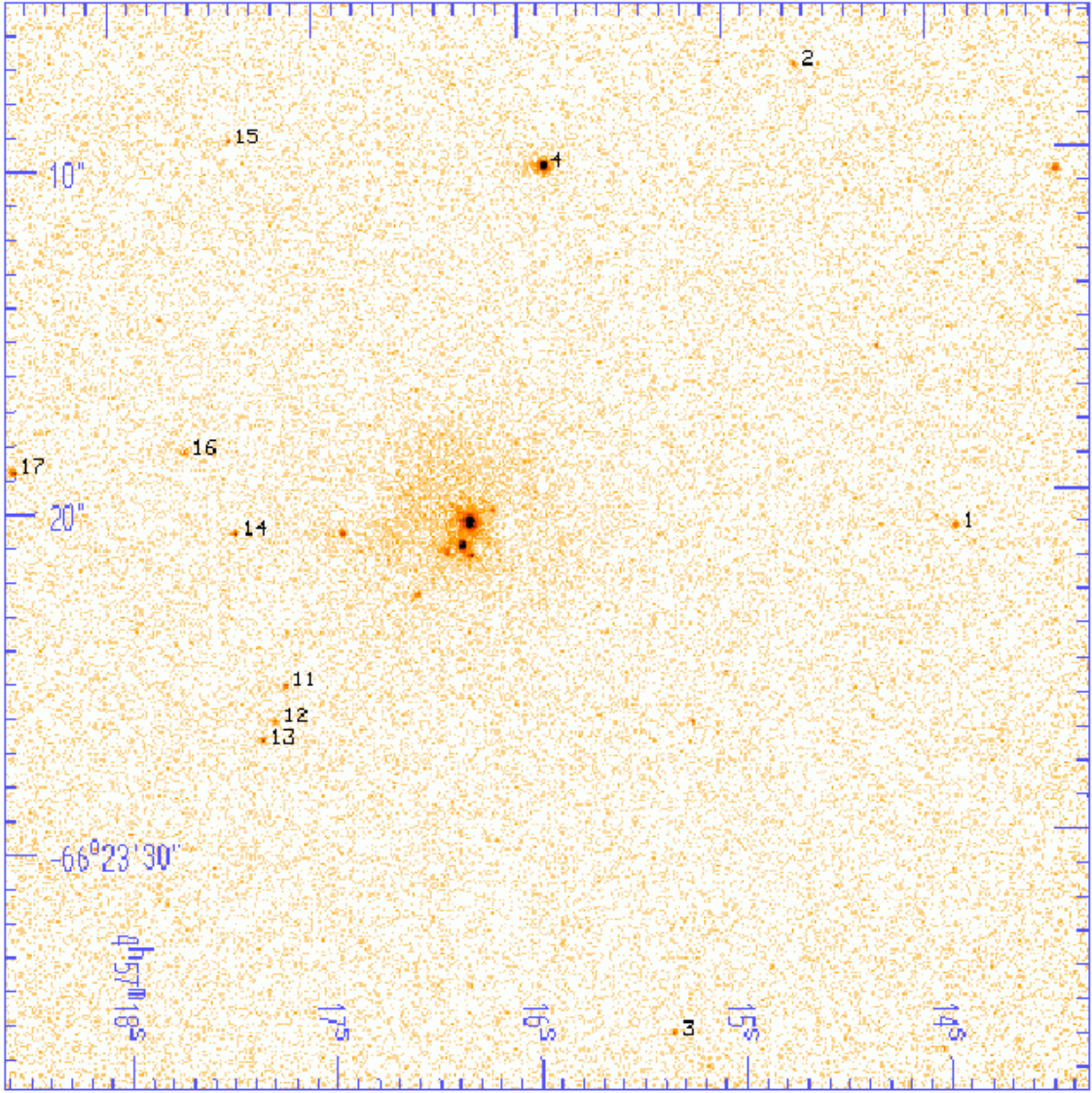}}
\caption{A WFPC2 image showing the LMC compact \h2\, N\,11A and its
adjacent field in the Str\"omgren $y$ filter (F547M) where the
brighter stars are labelled. A close-up image of the \h2\, region is
displayed in Fig.\,\ref{chart_core} and the photometry of the
stars is presented in Table\,\ref{phot}.  The field size is
\ab\,32\frac\,\x\,32\frac\, (\ab\,8\,pc\,\x\,8\,pc), and the
coordinates are in J2000. }
\end{center}
\label{chart_pc}
\end{figure*}

\begin{figure*}
\begin{center}
\resizebox{13cm}{!}{\includegraphics{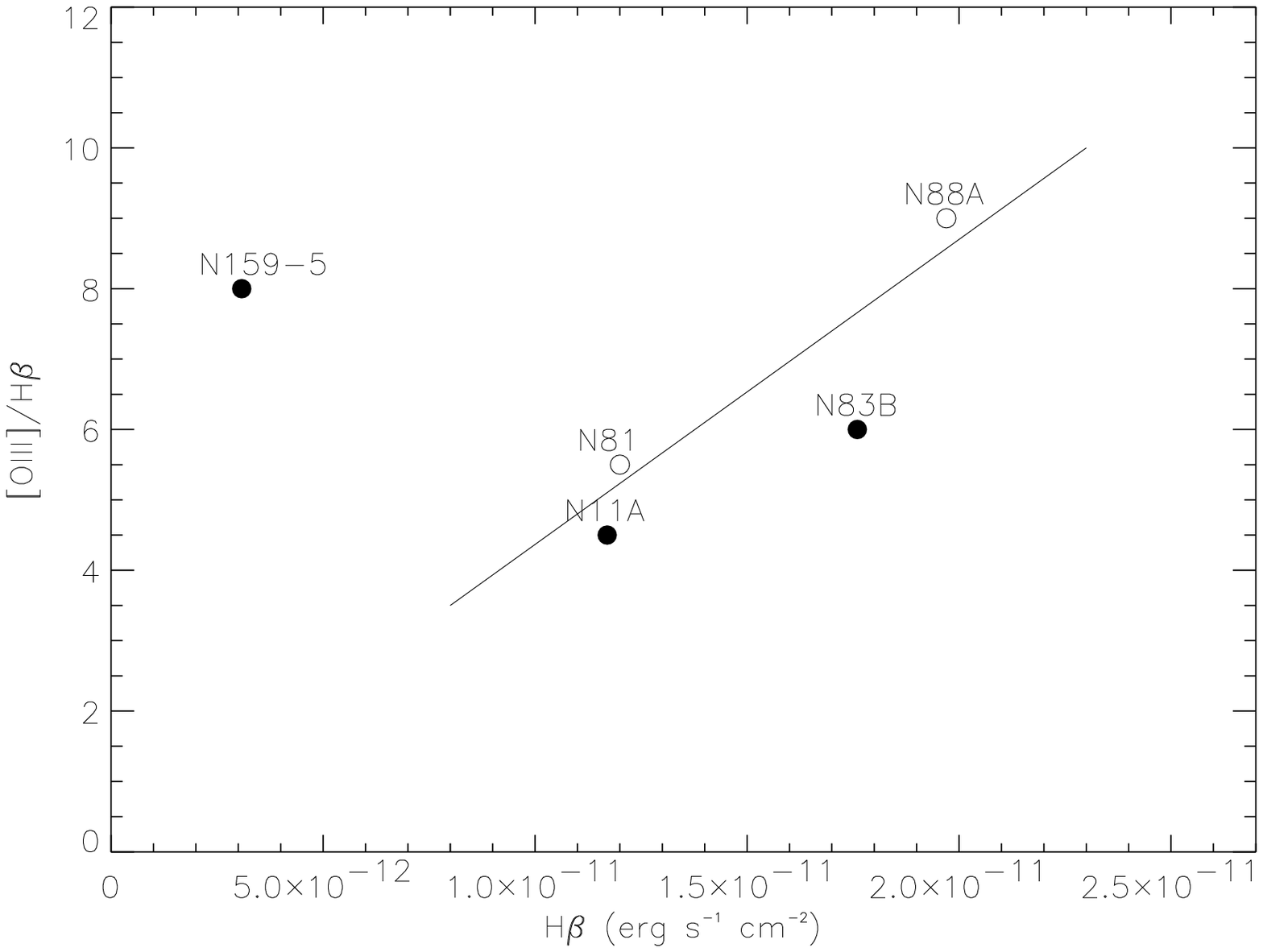}}
\caption{The variation of the \oiii\,(5007\AA)/\hb\, ratio for the Magellanic 
Clouds compact \h2\, regions with respect to their corresponding de-reddened
\hb\, flux. The full dots denote the LMC regions and the empty dots SMC 
ones. The line is simply indicative of the direction of the
corelation suggested by the models.}
\label{plot}
\end{center}
\end{figure*}

\section{Discussion}

The total \hb\, flux measured for N\,11A corresponds to a main exciting
source of spectral type O7.5--O8V. There are two major reasons, though,
which suggest that this estimate is a lower limit to the
power of the exciting star. One, as we mentioned earlier, is
the fact that the \h2\, region is most probably not fully
ionization-bounded and a fraction of the UV photons produced by the
star remain unaccounted for because they escape into the interstellar
medium.  A second factor which contributes to underestimating the
\hb\, flux is possibly an inadequate estimate of the extinction
correction, since the \hb\, flux was corrected assuming only external
extinction. Otherwise, it is also probable that the region harbors
hotter undetected stars still embedded in the gas/dust concentrations.
The same exciting star was observed by Parker et al. (\cite{par}) who
estimated its spectral type as O6-O3V. The O3 type,
though, is uncertain because of contamination by the strong nebular
emission line of \hei\,\lam\,4471\,\AA.  Moreover, if an O3V star lies
inside N\,11A, one would expect an \oiii/\hb\, ratio much higher than
that actually measured. \\

Ionization fronts in \h2\, regions, which represent the separation
between the hot ionized gas and the photodissociation region, can
appear differently according to their orientation with respect to the
line of sight. The high resolution {\it HST} images reveal a sharp
ridge north-east of the exciting stars which can be explained by the
presence of such a front. The associated molecular cloud in which the
stars of N\,11A formed must therefore be further to the north-east of
that front, in agreement with the observation that the dust content
increases significantly in that direction, as suggested by the higher
values of the Balmer decrement.  This result is supported by the
CO (1--0) observations of the N\,11 complex (Israel \& de Graauw
\cite{isr}, Caldwell \& Kutner
\cite{cal}).  The CO map shows that molecular emission comes from a
zone surrounding the central cavity of N\,11 and has a long branch
spreading in a north-east direction which passes by N\,11A. Although the
spatial resolution of the CO map is much coarser than that of the
present observations, there is little doubt about the relative
position of the molecular cloud with respect to N\,11A. \\

The morphology of N\,11A is well accounted for by the 
champagne model (Tenorio-Tagle \cite{teno}, Bodenheimer et
al. \cite{boden}). It corresponds to the evolutionary stage when 
the ionization front has reached the border of the molecular cloud.
The central bright area (Fig.\,\ref{true}) is the
cavity created in the molecular cloud by the ionizing photons of the
stars, and the bright ridge is a part of the cavity situated along the
line of sight.  Dense ionized gas is pouring out of the cavity into
the interstellar medium as shown by the presence of the outer diffuse
component.  Since the line of sight is slightly tilted with respect to
the outflow axis, a part of the cavity and the associated external
dust hides a portion of the \h2\, region and causes the higher
extinction observed at its north-east side. The surrounding
filamentary structure (Fig.\,3) is probably created by the stellar
winds from the central cluster acting on the dense ionized gas at the
surface of the molecular cloud.\\

The compact \h2\, region N\,11A is the fifth in a sample of HEBs we
have observed so far with  {\it HST}. The other regions of the sample are
N\,159-5 (the Papillon nebula) and N\,83B in the LMC (Heydari-Malayeri et
al. 1999c, 2001), as well as N\,81 and N\,88A in the SMC (Heydari-Malayeri
et al. 1999a, 1999b). Despite the small number of regions available, it
would be interesting to compare those HEBs by examining a possible
correlation between their de-reddened \hb\, fluxes and the
corresponding \oiii\,(5007\AA)/\hb\, ratios (Fig.\,6). The motivation
for that is due to  current photoionization models
of spherically symmetric \h2\, regions which suggest that the emission
line spectrum of the region depends on  the effective
temperature of the star(s), the ionization parameter, and the gas
metallicity (Stasi\'nska \cite{sta}, Stasi\'nska \& Leitherer
\cite{stalei}).  The ionization parameter is defined as $U=Q/4\pi
R^{2}nc$, where $Q$ is the total number of ionizing photons with
energy above 13.6 eV, $R$ the Str\"omgren radius, $n$ the hydrogen
density, and $c$ the speed of light.  Models by Stasi\'nska (1990)
show a linear correlation between
\oiii/\hb\, and the \hb\, flux for \h2\, regions with the same gas
density, the same number of exciting stars, but with increasing
effective temperatures. The reason for this relationship is that when
there are energetic UV photons available for producing high
\oiii\,/\hb\, ratios, they also produce a large \hb\, flux. 
The models also show that lower metallicity environments favor higher
\oiii/\hb\, ratios, but the metallicity dependence can be outweighed
by the ionization parameter.\\

Even though the real \h2\, regions are not ionization-bounded spheres,
and one should be careful when applying model results to observations,
the fact that HEBs have sizes which are of the same order of magnitude
provides a good opportunity for comparison. It is interesting that if
we exclude N\,159-5, the plot of Fig.\,6 suggests that there is indeed a
linear relationship as predicted by the models.  The reason for
discarding N\,159-5 is because we know (Heydari-Malayeri et al. 1999c)
that it is a very reddened region by both local and external dust, and
its discrepant position may be due to inadequate flux correction for
extinction in a manner similar to the one discussed at the beginning
of this section.  If our reasoning is justified, then the measured
\hb\, flux of N\,159-5 is underestimated by a factor of \ab\,5, 
mainly due to the effect of internal dust whose properties
remain to be studied. 
We also observe that among the HEBs studied,
SMC N\,88A is the most excited and LMC N\,11A the least excited.  The
lower excitation of N\,11A may simply be due to the fact that it is
excited by a less massive, young star. It may also be the result of an
evolutionary effect, in the sense that the lower \oiii/\hb\, may in
fact be the consequence of a smaller gas density caused by the
champagne outflow.  From our previous {\it HST} observations we know
that SMC N\,88A is younger than SMC N\,81, since it is more compact, of
higher gas density, and more affected by dust so that its exciting
stars are still hidden inside the \h2\, region. The same holds for the
LMC objects, since N\,159-5 is associated with a much larger amount of
dust than N\,83B and N\,11A and is more compact compared to them. \\
  
One should keep in mind though, that 
despite the compactness of N\,11A (and
HEBs in general) compared to typical extragalactic \h2\, regions, if it was
situated in our Galaxy, it would appear as a rather ``classical'' \h2\,
region   
nearly \ab\,4 times smaller than the Orion
nebula.  In fact, based on our new observation, it appears that N\,11A has
many similarities with Orion regarding its central exciting cluster, its
high excitation, a region of high extinction at the northern border
and the filamentary loops on the southern side. 
Consequently, these small high excitation, compact \h2\, regions found
in the Magellanic Clouds near locations where other giant \h2\,
regions are present, are very interesting to study in order to better
understand massive star formation processes in more distant
galaxies.

\begin{acknowledgements} 
We would like to thank the referee Dr. Joel Wm. Parker for his 
helpful comments. We are also grateful to Dr. Gra\.zyna Stasi\'nska 
for discussions.  VC would like to acknowledge the
financial support for this work provided by NASA through grant number
GO-8247 from the STScI, which is operated by the Association of
Universities for Research in Astronomy, Inc., under NASA contract
26555.
\end{acknowledgements}

{}

\end{document}